# Large Eddy Simulation of a NACA0015 Circulation Control Airfoil Using Synthetic Jets

**Pititat Itsariyapinyo**[*]**, Rajnish N. Sharma**

*Mechanical Engineering, University of Auckland, Auckland, 1142, New Zealand*

## Abstract

Large eddy simulation of a NACA0015 circulation control airfoil using synthetic jets is conducted. A chord Reynolds number of $1.10 \times 10^5$, the excitation frequency of 175 Hz, the momentum coefficients of 0.0044 to 0.0688, and the angles of attack of 0°, 6°, and 12° are employed in this study. The numerical results presented in this paper show good agreement with the experimental results. Fluctuations in the lift and drag coefficients on the non-actuated airfoil are found to be governed by the vortex shedding frequency. Provided that the momentum coefficient is sufficiently high, the lift and drag coefficients tend to fluctuate at the synthetic jet frequency. The mean lift coefficient linearly increases with the momentum coefficient at a low momentum coefficient range and its incremental rate begins to decline at a high momentum coefficient range. Increasing the angle of attack of the airfoil is observed to slightly reduce the slope of the lift increment.

## 1 Introduction

The past decades have witnessed much development in the circulation control airfoil. The concept of a circulation control airfoil first emerged in the late 1930s when forcing an air jet on an airfoil surface was found to increase the lift coefficient [1]. Followed by this finding, circulation control technology was demonstrated on a Hunting H.126 aircraft as an alternative high-lift device in 1963 [2]. In 1974, researchers at West Virginia University installed a circulation control wing on a BD-4 homebuilt aircraft and successfully conducted flight tests [3],[4] which prompted them another project by the Navy and Grumman Aerospace to convert A-6A bombers for short take-off and landing (STOL) operation [5],[6]. A recent appearance of a circulation control airfoil could be seen in a DAEMON UAV aircraft which uses circulation control technology for flight control [7].

Conventionally, a circulation control airfoil consists of an air plenum, an orifice, and a rounded trailing edge (Coanda surface) which is designed to facilitate the Coanda effect. Compressed air is usually either supplied from a compressor or a propulsive system to produce an air jet through an orifice [8]. Provided that the angle of attack of an airfoil is below the stall angle and that the flow does not completely separate from an airfoil, flow separation on this airfoil would probably occur somewhere around the Coanda surface. On the Coanda surface of an unforced circulation control airfoil (non-actuated airfoil), the shear flow on such airfoil, which generally does not have sufficient momentum to surmount the curvature of the Coanda surface, usually separates as soon as it reaches the Coanda surface. Upon forcing an air jet onto the Coanda surface, additional momentum offered by an air jet usually reenergizes the local shear flow which enables it to remain attached and entrained on the Coanda surface for a longer angular distance. Through the process of promoting momentum transfer within the shear flow, flow separation on this airfoil could be delayed [9]. Provided that a high-momentum jet is employed and a flow separation point is pushed further downstream, the lift coefficient usually increases in response to an improved circulation on an airfoil [10].

Although a circulation control airfoil has enjoyed decades of development, its ability to effectively exploit vortex shedding is somewhat questionable. While a high-momentum air jet could easily be produced by either a compressor or a propulsive system, the nature of a continuous jet, which is produced by these sources, lacks the oscillatory behavior which is usually exhibited by vortex shedding [11]. Thus, additional moving components are commonly introduced to the system to pulse an air jet. Alternatively, one may opt to use a synthetic jet actuator to produce successive suction and blowing at a desired frequency. As opposed to a conventional circulation control

---

[*] Corresponding author.

E-mail addresses: pits055@aucklanduni.ac.nz (P. Itsariyapinyo), r.sharma@auckland.ac.nz (R.N. Sharma).

airfoil which relies on a plumbing system, a synthetic jet actuator consists of an orifice, oscillating membrane, and a cavity. Since an oscillating membrane, such as a loudspeaker and a piezoelectric membrane, is normally driven by a series of sinusoidal waves, successive suction and blowing are produced by the ingestion and expulsion strokes which take place when an oscillating membrane oscillates outward and inward, respectively. Hence, a synthetic jet actuator is also known as a "zero-net-mass-flux" (ZNMF) device.

Knowing that vortex shedding is periodically recurring phenomenon which is responsible for fluctuations in aerodynamic forces [12], the effectiveness of an oscillatory excitation is largely dependent on the choices of the excitation frequency, momentum coefficient, and the excitation location [13]. In many investigations, excitations at the Strouhal number which is on an order of the vortex shedding frequency are found to yield satisfactory results [14]. In the previous study which concerned the wind tunnel testing of a NACA0015 circulation control airfoil [15], the diameter of the rounded trailing edge on the circulation control airfoil is assigned as its characteristic length as such dimension is likely to be responsible for the wake width of this airfoil. The vortex shedding frequency is estimated from the spectral analysis of wake velocities at $x/c = 2.10$ (vertical-component or y-direction fluctuating velocities, $v$) to be approximately 0.17 when the flow is fully attached on the airfoil. The most effective excitation frequency, which is capable of yielding the highest lift coefficient, is corresponding to the Strouhal number of 0.14. Similar to the aforementioned investigations, the results from the experimental study seem to agree well with the others in which the most effective excitation frequency is found near the vortex shedding frequency [8].

The objectives of the current study are to validate the numerical results with the experimental results and attempt to explore some other aspects of a circulation control airfoil via flow visualization which tend to be more difficult to do so in wind tunnel testing. Since the effects of excitation frequency on the aerodynamic forces have been investigated to a certain extent in the previous study [15], the current study will focus more on the effects of momentum coefficient as the synthetic jet actuators used in the experimental study provide relatively low momentum coefficient. Thus, the excitation frequency of 175 Hz ($Sr = 0.14$), the momentum coefficients of 0.0044 to 0.0688, and the angles of attack of 0°, 6°, and 12° are employed in this numerical study.

## 2 Numerical Setup

In addition to the effectiveness of a circulation control operation from the perspective of flow control, the geometrical parameters of a circulation control airfoil are also responsible for the characteristics of jet turning [16],[17]. According to Englar's hypothesis [18], the ratios of a jet height to the Coanda radius ($b/r$), the jet height to the airfoil chord length ($b/c$), and the Coanda radius to the airfoil chord length ($r/c$) of 0.05, 0.0012, and 0.0235 were chosen to ensure effective circulation control operation, respectively. The designations of these geometrical parameters are as shown in Figure 1a and c. For the sake of consistency between the numerical and experimental studies, the airfoil chord length of 170 mm was used. Furthermore, a free-stream velocity of 10 m/s was assigned to the free-stream inlet to maintain the chord Reynolds number of $1.1 \times 10^5$. The geometries and meshes of the flow domain were created in ICEM CFD 15.0 using an unstructured hexahedral mesh. Pre-CFX, CFX Solver, and Post-CFD of ANSYS CFD R15.0 Academic edition were used to produce numerical results.

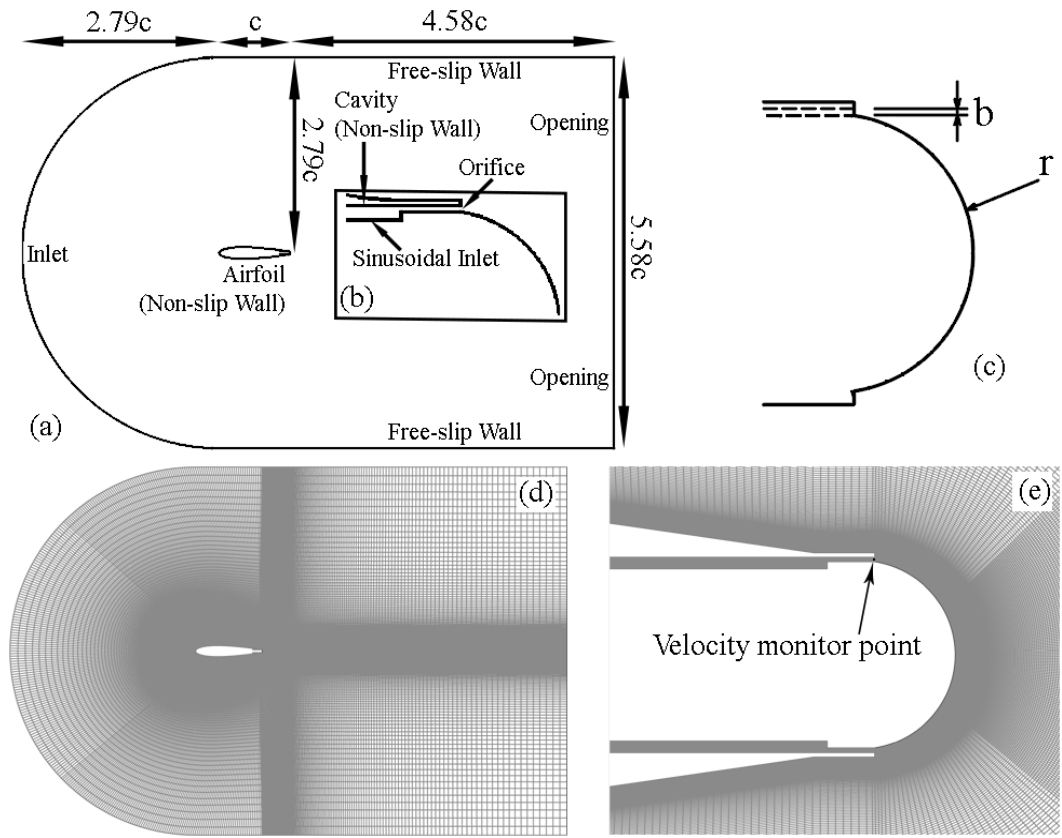

**Figure 1 Boundary conditions of (a) a flow domain and (b) the synthetic jet actuator, (c) the schematics of the Coanda surface, and the meshing topology around (d) the airfoil and (e) the rounded trailing edge.**

The C-grid configuration, as shown in Figure 1a, b, and d, was employed for this simulation. The round arc with a radius of $2.79c$ was assigned as an inlet and prescribed with a free-stream velocity. The vertical distance from the top to bottom walls of the flow domain was $5.58c$. The top and bottom walls were assigned free-slip walls. The streamwise distance from the trailing edge to the far-end wall was $4.58c$. The far-right wall was assigned as an opening. The surfaces of the airfoil, orifice, and the cavity were assigned as non-slip walls. The side walls of this flow domain were assigned as symmetry planes. The airfoil span length of $0.07c$ was used in the simulation as it was found to be capable of resolving the turbulent coherence structure on an airfoil while optimizing the computational time [19]]. It is noted that the size of this flow domain is based on similar numerical studies which were able to yield satisfactory results while optimizing the computational time [20]. To put the flow domain size into perspective, 4 different flow domains with the surface areas of 1.26 (used for the current study), 2.01, 2.85, and 5.03 $m^2$ were tested prior to the finalization of the flow domain size. In this analysis, differences in the aerodynamic coefficients and the RMS fluctuating velocities predicted by the smallest and the largest flow domains are typically 2-4%. Due to the fact that the largest flow domain takes approximately 218 hours to solve 10000 time steps while the smallest flow domain only takes approximately 145 hours to solve the same number of time steps, the smallest flow domain is used in the current study as further enlarging the flow domain size has very little effect on the results. The oscillating membrane was modelled as a sinusoidal velocity inlet [21] whose amplitude is defined by the form $U_j(t) = U_d \sin(2\pi f t)$ where $U_d$ is the membrane velocity and $f$ is the excitation frequency. The value of the synthetic jet velocity at the orifice ($U_j$), which was used to estimate the momentum coefficient, was obtained from the monitor point which is placed at the orifice exit (see Figure 1e). In order to be able to resolve the near-wall flow, the $y^+$ value for the first near-wall node was kept slightly below 1.

Since vortex shedding is reportedly responsible for fluctuations in lift coefficient [22]-[24], it is important that the Courant number of less than or equal to 0.04 was used to determine the appropriate time step which could satisfy the Courant-Friedrichs-Lewy (CFL) condition [25] and resolve the timescale of vortex shedding on the airfoil. Thus, the time step used for all simulation cases was 1/(175×250) or approximately 2.29×10⁻⁵s. For all the simulation cases, 5000 time steps were solved and employed to produce the transient and time-averaged results as the convergence was usually achieved at approximately 3000 time steps where the aerodynamic forces begin to exhibit quasi-periodic trends. For the few selected simulation cases where the spectral analysis is performed to reveal the dominant frequency, 30000 time steps were solved and employed to produce high-resolution frequency spectra.

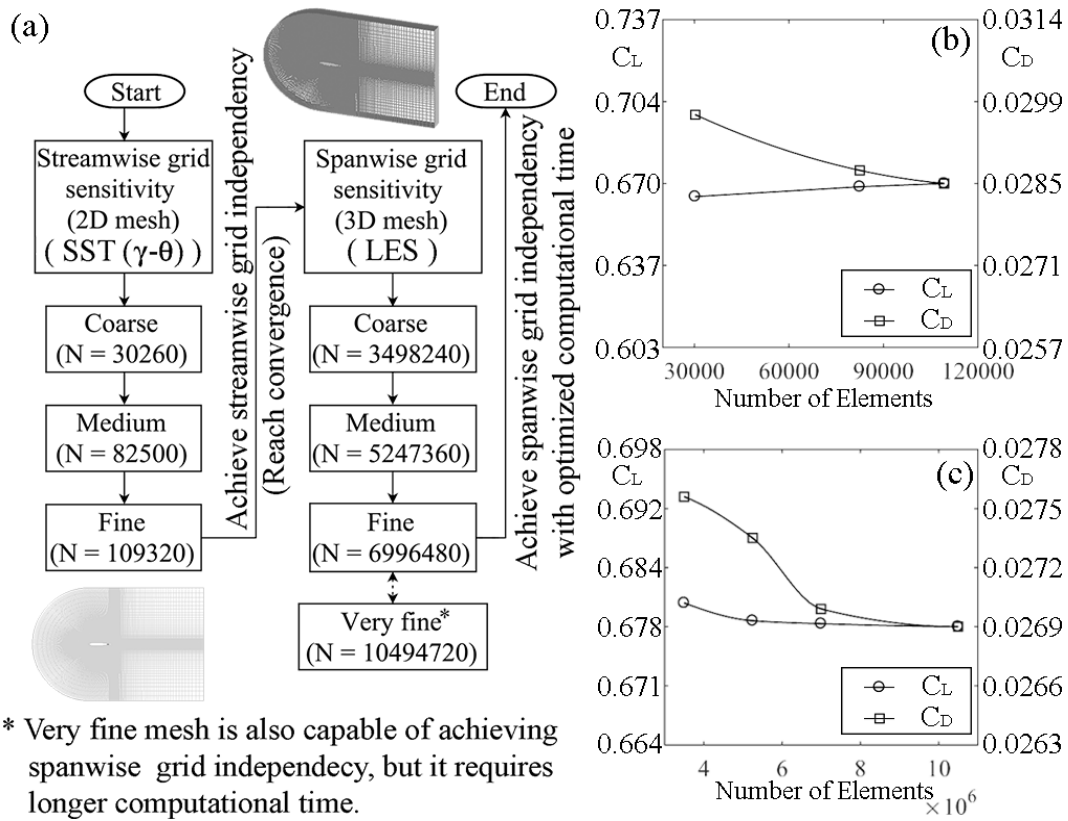

**Figure 2 (a) Flow chart describing the procedures taken to perform the streamwise and spanwise grid sensitivity analyses and typical plots of (b) streamwise and (c) spanwise grid sensitivity analyses obtained from the airfoil at α = 6°.**

Prior to conducting large eddy simulations (LES), the streamwise and spanwise grid sensitivity analyses were carried out to optimize the number of mesh elements required to achieve convergence. For the streamwise grid sensitivity analysis, URANS SST (γ-θ) model (Langtry & Menter Transition Model (γ-θ)), high resolution advection scheme, and second order backward Euler transient scheme were used to solve pseudo three-dimensional simulations whose flow domains were extruded in the spanwise direction by 1 mesh element. Among 3 different mesh qualities (coarse, medium, and fine), the fine mesh was found to achieve streamwise grid independency. For the spanwise grid sensitivity analysis, LES (WALES), central difference advection scheme, and second order backward Euler transient scheme were used to solve three-dimensional simulations. For the purpose of determining the number of spanwise mesh elements which could achieve spanwise grid independency and optimize the computational time, the pseudo three-dimensional fine mesh, which was acquired from the streamwise grid sensitivity analysis, was extruded in the spanwise direction by 32 (coarse), 48 (medium), 64 (fine), and 96 mesh elements (very fine). In this analysis, it was found that the aerodynamic forces and the timescale of the vortex shedding of the fine mesh were very similar to those of the very fine mesh. Thus, the fine mesh of the spanwise grid sensitivity analysis was employed to generate numerical results as such configuration required less computational time. The mesh configurations and turbulence models used in the simulation and typical plots of streamwise and spanwise grid sensitivity analyses were shown in Figure 2a, b, and c, respectively. For reference, 6 computational nodes with 96 CPUs can solve 10000 time steps of the coarse, medium, fine, and very fine meshes within approximately 40, 75, 145, and 275 hours, respectively.

In the model validation section, the experimental and numerical two-dimensional (2D) lift ($C_l$) and drag ($C_d$) coefficients were acquired from pressure distributions at the mid-span of the airfoil. In the experimental study, static pressure taps on the airfoil were employed to collect the mean surface pressures at the mid-span of the airfoil. These mean surface pressures were obtained by means of time-averaging. An integration of the mean pressure coefficients acting on the airfoil was carried out to compute the lift and drag coefficients. The uncertainties ($2\sigma$) of the experimental lift and drag coefficients are typically below 2% and 5%, respectively. Due to the limitation of the traverse system in the experimental setup which limits the traversing distance of a TSI cobra probe (series 100), fluctuations in the experimental y-direction velocities ($v$) were measured at $x/c = 2.10$ and at the mid-span of the airfoil. In the numerical study, a polyline was created at the mid-span of the airfoil surface to acquire the mean surface pressures which were then reduced to lift and drag coefficients. Like the experimental results, these mean surface pressures were also obtained by means of time-averaging. For the flow domain used in this study, fluctuations in the numerical y-direction velocities were measured at $x/c = 1.10$ and at the mid-span of the airfoil as the flow in a denser mesh region (near to the airfoil wall) tends to be better resolved. It is noted that this difference in the measurement location does not affect the value of the vortex shedding frequency as such frequency is found to be constant from at least x/c = 1.00 – 2.80 [26]. While these setups may give rise to some discrepancies in the velocity magnitude between the experimental and numerical studies, the objective of the spectral analysis in this study is to mainly compare the dominant frequencies obtained from the experimental and numerical studies.

In the results section, the numerical three-dimensional (3D) lift ($C_L$) and drag ($C_D$) coefficients were acquired from a force function provided by the CEL expression in ANSYS CFX. By doing so, the contribution of viscous stress to the aerodynamic forces can be taken into account. It is also important to note that the data presented in the results section were taken from LES cases as those obtained from URANS cases were conducted to perform streamwise grid sensitivity analysis only.

Since a laminar separation bubble, which is indicated by a "kink" on the pressure distribution, is enclosed by a flow separation point and a flow reattachment point [27], a flow separation point is usually found at the point where the pressure coefficient refuses to follow the pressure recovery trend whereas a flow reattachment point is found at the point where the pressure coefficient continues to follow the pressure recovery trend. In this study, the streamwise distances or the lengths of laminar separation bubbles in the experimental study were estimated from the linear fits to the pressure distributions at the upstream and downstream regions of a laminar separation bubble for a flow separation point and a flow reattachment point, respectively [28]. The estimated uncertainty of these flow separation and reattachment points is typically within 2% of the chord length. For the sake of consistency, the lengths of laminar separation bubbles in the numerical and experimental studies are estimated in the same way when the model validation is carried out.

In this study, the vortex shedding and synthetic jet frequencies are normalized by the diameter of the Coanda surface $d$ and a free-stream velocity $U_\infty$ to yield the Strouhal number $Sr$.

$$Sr = \frac{fd}{U_\infty} \tag{1}$$

In an attempt to establish a measure for the synthetic jet momentum so that the numerical results could be validated with the experimental results, the formula for the synthetic jet momentum coefficient is expressed as

$$C_\mu = \frac{\bar{I}_j}{1/2\,\rho_\infty U_\infty^2 c} \tag{2}$$

$$\bar{I}_j = \frac{1}{T/2}\rho_j b \int_0^{T/2} U_j^2(t)\,dt \tag{3}$$

where $\bar{I}_j$ is the time-averaged jet momentum per unit length during the expulsion stroke, $T$ is the actuation period or the reciprocal of the actuation frequency $f$, $\rho_j$ is the air density, $U_j$ is the jet velocity at the exit of the slot of the actuator, $b$ is the width or height of the jet slot, $\rho_\infty$ is the free-stream density, $U_\infty$ is a free-stream velocity, and $c$ is the chord length.

# 3 Model Validation

Prior to the presentation and discussions of the numerical results, it is of great importance that some numerical results are validated by the experimental results. In this case, the time-averaged pressure distributions and the aerodynamic forces obtained from the experimental results at $\alpha$ = 0°, 6°, and 12° are provided to complete the task.

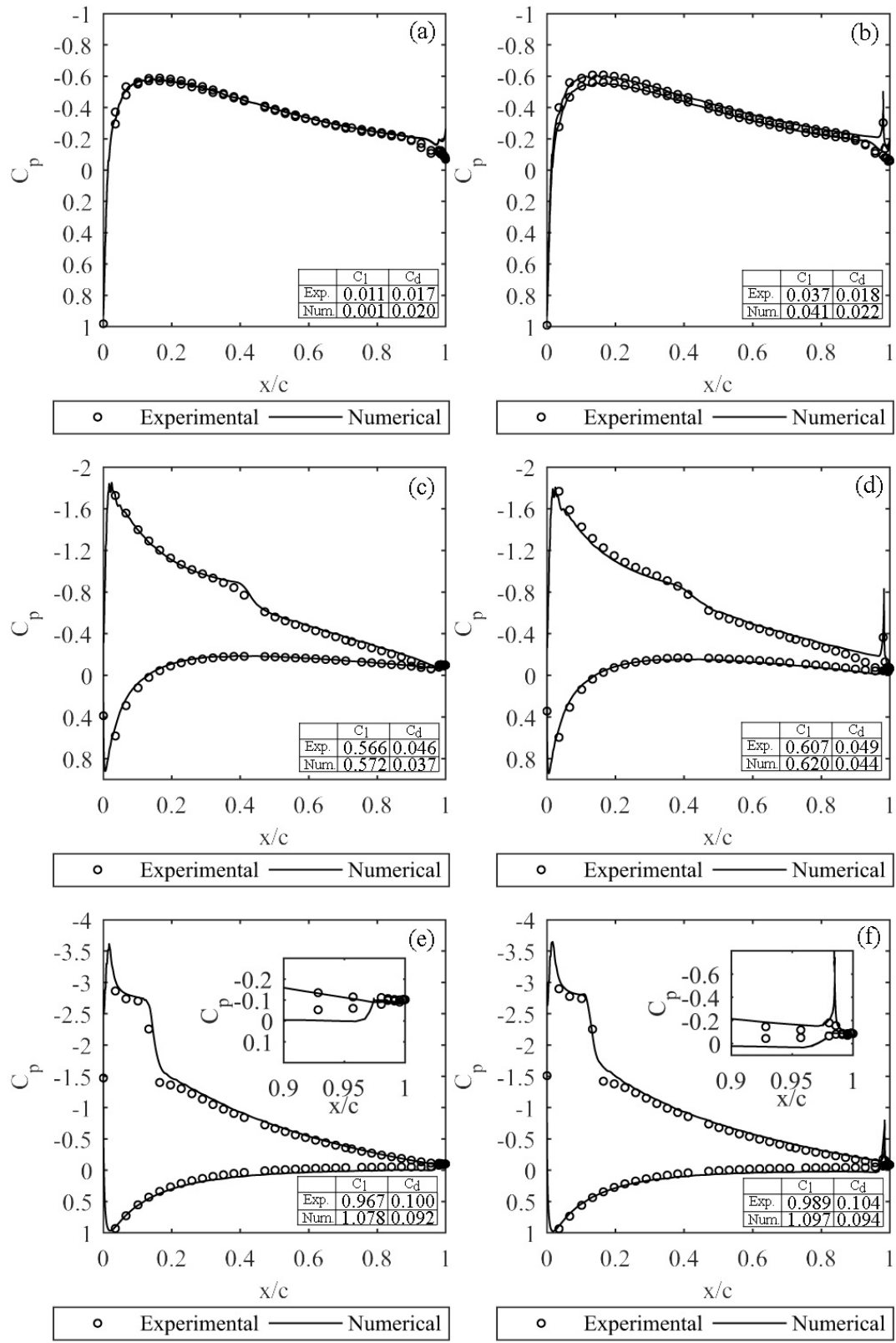

**Figure 3** Pressure distributions of experimental and numerical studies on the non-actuated airfoil at $\alpha$ = (a) 0°, (c) 6°, and (e) 12° and the actuated airfoil ($C_\mu$ = 0.0044) at $\alpha$ = (b) 0°, (d) 6°, and (f) 12°.

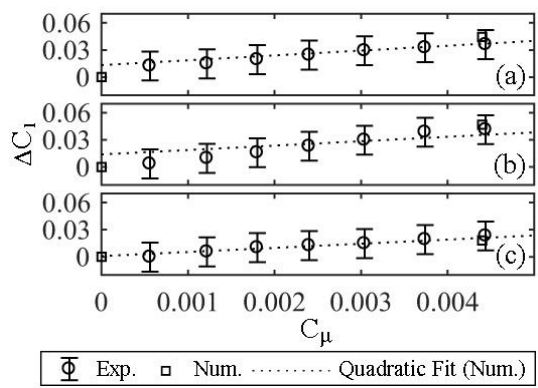

**Figure 4 Comparison between changes in the experimental and numerical 2D lift coefficients ($C_l$) at α = (a) 0°, (b) 6°, and (c) 12°.**

As shown in Figure 3, the pressure distributions obtained from the numerical study conform well to those of the experimental study, especially at α = 0° and 6° where the appearances of laminar separation bubbles are subtle. The numerical 2D aerodynamic forces of both non-actuated and actuated cases agree well with the experimental 2D aerodynamic forces. Even though the percentage differences between the experimental and numerical studies may appear to be somewhat high in some cases, it is noteworthy that such is usually due to the fact that the aerodynamic forces of a pre-stall airfoil are typically low. Thus, the percentage differences between the experimental and numerical results are very sensitive to small differences in the absolute values. The discrepancies between the 2D aerodynamic forces of the non-actuated and actuated cases become slightly more apparent at α = 12°; however, changes in the experimental and numerical 2D lift coefficients at all angles of attack, as shown in Figure 4, are still in good agreement.  The differences between the experimental and numerical results are within 1-12% for 2D lift coefficient. The flow separation and reattachment points on the non-actuated airfoil at α = 6° are found at $x/c$ = 0.252 and 0.447 for the experimental study and $x/c$ = 0.267 and 0.429 for the numerical study. On the other hand, the flow separation and reattachment points on the actuated airfoil ($C_\mu$ = 0.0044) at α = 6° are found at $x/c$ = 0.381 and 0.441 for the experimental study and $x/c$ = 0.377 and 0.416 for the numerical study. Since there are not enough pressure taps on the leading edge of the airfoil used in the experimental study to estimate the location of a flow separation point on the airfoil at α = 12°, only the flow reattachment points are estimated. The flow reattachment points on the non-actuated and actuated airfoils ($C_\mu$ = 0.0044) at α = 12° are found at $x/c$ = 0.158 and 0.150 for the experimental study and $x/c$ = 0.159 and 0.143 for the numerical study, respectively. Therefore, the differences between the locations of the flow separation and reattachment points obtained from the experimental and numerical studies are within 1-6%.

## 4   Results and Discussion

Although a considerable amount of knowledge on a circulation control airfoil was obtained from the previous experimental study [15], there are some aspects which pose great difficulties when such are tackled by an experimental approach. For instance, the synthetic jet actuator used in the experimental study provides relatively low synthetic jet velocity or low momentum coefficient. As a consequence, the experimental investigation on the effects of momentum coefficient is limited to a narrow range of momentum coefficient.

Due to the limitations of the experimental approach, a numerical approach is employed in this study to explore the effects of vortex shedding and synthetic jet actuation on lift and drag coefficients. Since it was found in the experimental study that the diameter of the Coanda surface on the circulation control airfoil is likely to be associated with the vortex shedding frequency and that such frequency is reportedly influencing fluctuations in both lift and drag profiles, it is of interest that the effects of vortex shedding frequency on the lift and drag profiles are numerically and visually expressed. Similar to vortex shedding which recurrently takes place on an airfoil as a result of Kelvin-Helmholtz excitation, an oscillatory excitation offered by a synthetic jet could be similarly seen as a perturbation on the flow which is intentionally introduced to an airfoil to compensate for the momentum which may have been lost to vortex shedding. Due to this reason, the effects of synthetic jet actuation on the lift and drag profiles are investigated as well. Followed by an investigation on the characteristics of fluctuating lift and drag coefficients, the effects of synthetic jet actuation on the mean lift coefficients are studied to explore the relationship between the mean lift coefficient and the momentum coefficient.

## 4.1 Fluctuations in Lift and Drag Coefficients of Baseline Cases

Any prior knowledge on the dominant frequency of a wetted object, which typically represents the vortex shedding frequency, usually offers some insights into the effective flow control via an oscillatory excitation. In many investigations, it has been found that the lift coefficients of a bluff body are fluctuated at the vortex shedding frequency [22]-[24] and that the vortex shedding frequency could be empirically determined by the wake size of a bluff body [29]. In the case of the other arbitrary shapes, an appropriate characteristic length must be sought to estimate a definitive length scale of its wake size. In the experimental study, the diameter of a round-trailing-edge airfoil at $\alpha = 0°$ yields the Strouhal number of 0.17 (corresponding to $f = 210$ Hz) which scales relatively well to the universal Strouhal number ($Sr \approx 0.20$) [29].

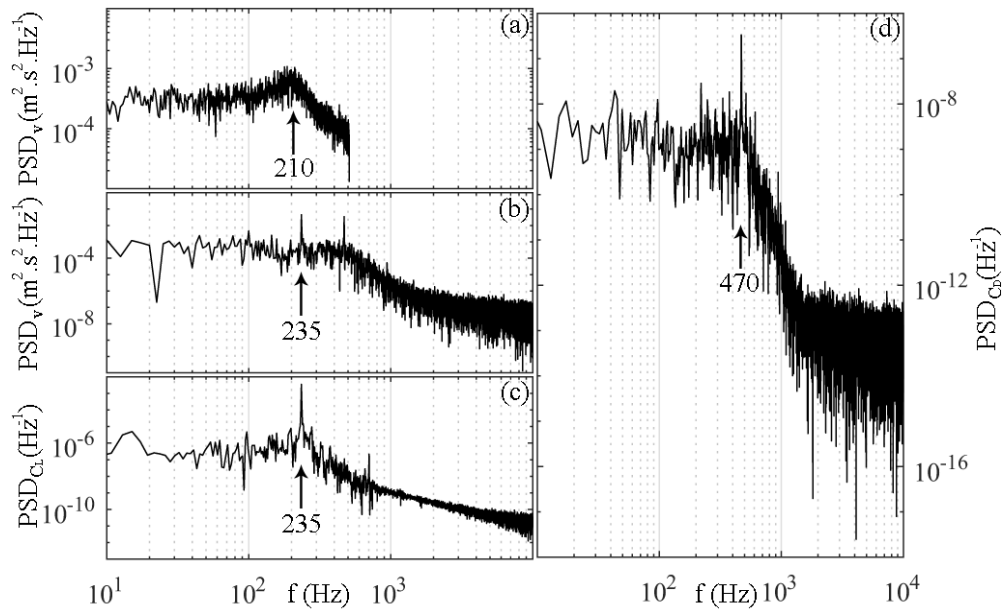

**Figure 5 Spectra of (a) fluctuating experimental y-direction velocities, (b) fluctuating numerical y-direction velocities, (c) fluctuating numerical lift coefficients, and (d) fluctuating numerical drag coefficients on the airfoil at $\alpha = 0°$.**

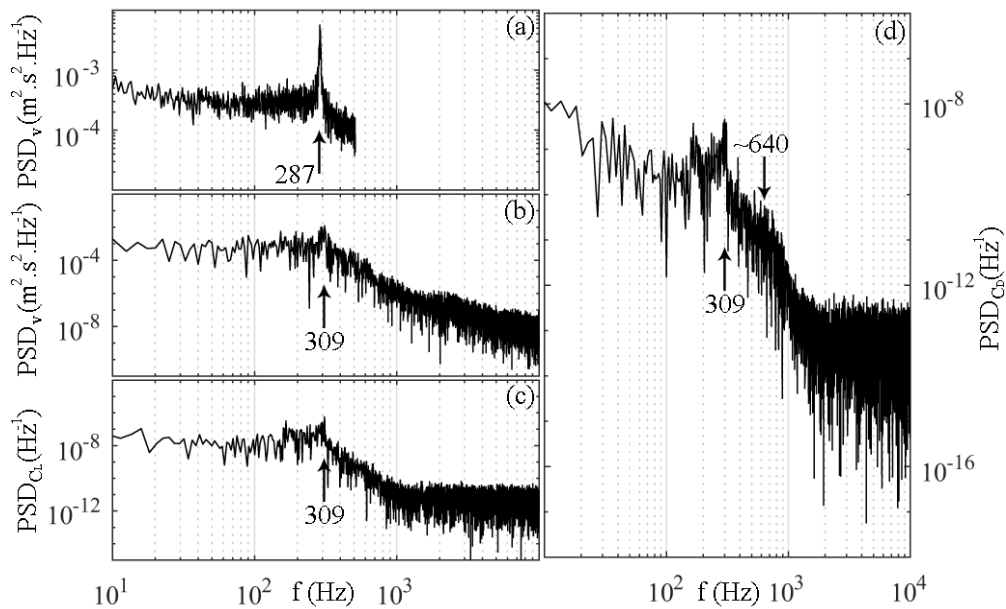

**Figure 6 Spectra of (a) fluctuating experimental y-direction velocities, (b) fluctuating numerical y-direction velocities, (c) fluctuating numerical lift coefficients, and (d) fluctuating numerical drag coefficients on the airfoil at $\alpha = 6°$.**

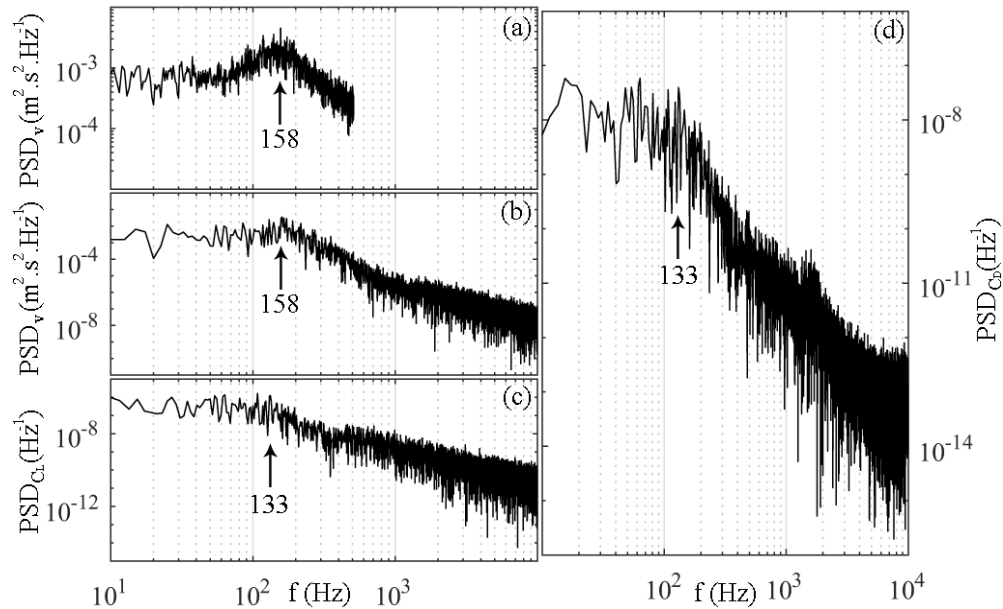

**Figure 7 Spectra of (a) fluctuating experimental y-direction velocities, (b) fluctuating numerical y-direction velocities, (c) fluctuating numerical lift coefficients, and (d) fluctuating numerical drag coefficients on the airfoil at α = 12°.**

Figure 5, Figure 6, and Figure 7 show the spectra of fluctuating experimental y-direction velocities, fluctuating numerical y-direction velocities, fluctuating numerical lift coefficients, and fluctuating numerical drag coefficients on the airfoil at $\alpha = 0°$, 6°, and 12°, respectively. The differences between the experimental and numerical vortex shedding frequencies which are obtained from y-direction velocities are less than 12%. Due to a technical limitation mentioned in the numerical setup section, differences in the magnitudes of the power spectral density between the experimental and numerical studies are expected. Nevertheless, it is noted that the main objective of this spectral analysis is to compare the dominant frequencies between those of the experimental and numerical studies. In order to reassure that a series of lift coefficients actually fluctuates at or close to the vortex shedding frequency, the spectra of fluctuating lift coefficients are plotted against those of y-direction velocities. The results show good agreement between the fluctuating lift frequencies and the vortex shedding frequencies for all angles of attack in which the fluctuating lift frequencies are identical or similar to the vortex shedding frequencies. Similar to other investigations on a circular cylinder [22]-[24], it is evident from the spectra of fluctuating drag coefficients at $\alpha = 0°$ (Figure 5d) that a series of drag coefficients fluctuates at twice the fluctuating lift frequency. However, at $\alpha = 6°$ and 12°, the fluctuating drag frequencies (Figure 6d and Figure 7d) are equal or similar to their respective vortex shedding frequencies.

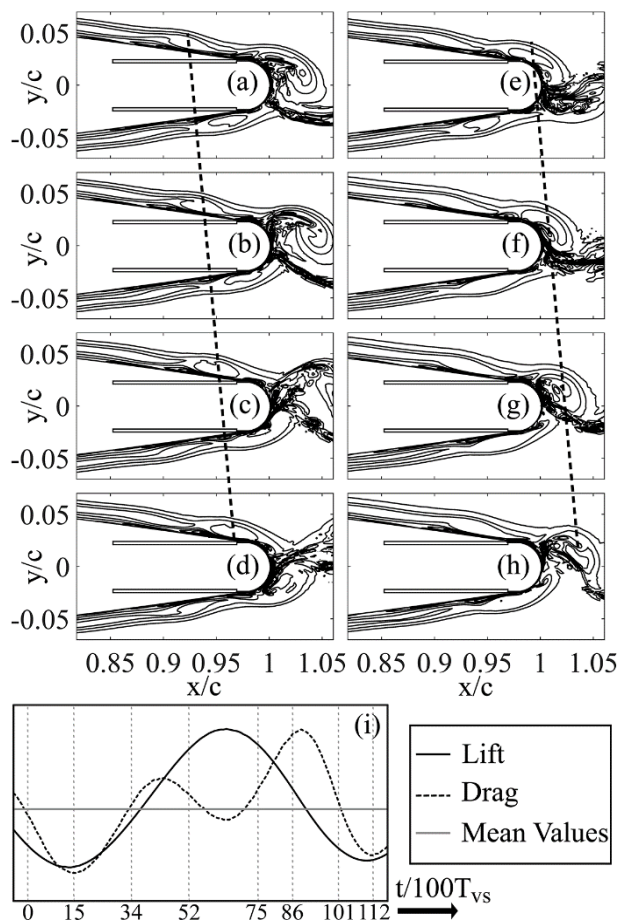

**Figure 8 Sequence of vortex shedding as illustrated by contour maps of spanwise vorticity at t/T$_{vs}$ = (a) 0, (b) 0.15, (c) 0.34, (d) 0.52, (e) 0.75, (f) 0.86, (g) 1.01, and (h) 1.12 and (i) series of fluctuating lift and drag coefficients on the non-actuated airfoil at α = 0°. Dash lines are overlaid on the sequence to trace the propagating vortex. T$_{vs}$ is a period of vortex shedding.**

Figure 8a-h illustrate the trajectory of vortices on the non-actuated airfoil at α = 0° with respect to a cycle of vortex shedding ($T_{vs}$) while Figure 8i exhibits the corresponding time-history results of lift and drag coefficients as these vortices propagate through different locations on the airfoil. Since vortices on the airfoil could easily be visualized by the spanwise vorticity when there is very little breakdown in the vortical structures or the spanwise variations, the airfoil at α = 0° is chosen to visualize vortex shedding which is also commonly found at other angles of attack. At $t/T_{vs} = 0$ where the lift and drag coefficeints move toward the minimum values or the trough, one of the vortices on the suction side is about to shed from the trailing edge. At $t/T_{vs} = 0.15$, the lift and drag coefficients reach their minimum values by the time the vortex on the suction side leaves the trailing edge. While vortex shedding on the suction side is responsible for the trough of flcutuating lift coefficients, the maximum lift coefficient is achieved when the vortex on the pressure side shed from the traling edge at $t/T_{vs} \approx 0.64$. Though the flow visualization at this particular is not shown, it is evident from $t/T_{vs} = 0.52$ and 0.75 that vortex shedding on the pressure side commences within these timeframes. On the other hand, the drag coefficient is observed to rise and fall to the maximum and minimum when the vortex on either suction and pressure side arrives at and lifts off from the trailing edge, respectively. Thus, drag coefficient is fluctuating at twice the fluctuating lift frequency. At $0.34 < t/T_{vs} < 0.52$, the vortex on the pressure side is responsible for the uphill and the downhill of fluctuating drag coefficients as this vortex approaches and lifts off from the trailing edge, respectively. Similarly, the uphill and downhill trips of fluctuating drag coefficients are illustrated in a more gradual manner at $0.75 < t/T_{vs} < 1.01$. Within these timeframes, the drag coefficient is at the mean value on the uphill side, the peak, and the mean value on the downhill side when the vortex on the suction side approaches, arrives at, and lifts off from the trailing edge, respectively. From $t/T_{vs} = 0.86$ to 1.12, the next cycle of vortex shedding is underway.

So far, it has been found that 2 vortices are shed from the airfoil within a cycle of vortex shedding. Vortices on the suction side are responsible for the trough of lift fluctuations while those on the pressure side are responsible for the peak of the fluctuations. Since a cycle of lift fluctuations is corresponding to a cycle of vortex shedding, it is fair to say that the lift coefficient on the airfoil fluctuates at the vortex shedding frequency which was also found to be true in the case of a circular cylinder [22]-[24]. In terms of the value of the vortex shedding frequency, the period in between $0.37t/T_{vs}$ and $0.89t/T_{vs}$, which marks half cycle of vortex shedding, is approximately $2.24 \times 10^{-3}$s. Thus, the period of a full cycle of vortex shedding is approximately $4.48 \times 10^{-3}$s which is corresponding to 223 Hz. Since the spectra are obtained from thousands of samples whereas the one presented here is acquired from a single cycle of vortex shedding, a slight difference between the 2 values is expected. Nevertheless, the spectra of lift and drag coefficients (Figure 5b) are indeed reflected by the fluctuations of lift and drag coefficients which take place within a cycle of vortex shedding (Figure 8). Even though the main concern of the current study is the estimation of the vortex shedding frequency, the period of $3.33 \times 10^{-3}$s, corresponding to 300 Hz, is reported by an exeprimental study to be the shear-layer instability frequency [15]. Interestingly, this period of time is corresponding to the the frequency at which a vortex rolls up. As shown in Figure 8a-e , it is noticable that 2 cat's eyes are produced within $0 < t/T_{vs} < 0.75$ and are visible on the suction side at $0.90 < x/c < 0.95$. The period between these 2 timeframes is $3.20 \times 10^{-3}$s which is equated to approximately 313 Hz.

## 4.2 Fluctuations in Lift and Drag Coefficients of Actuated Cases

Similar to a spectral peak of the vortex shedding mode which is most likely to be produced by the excitation of Kelvin-Helmholtz instability, a spectral peak which represents the synthetic jet actuation could possibly be produced by the excitation of a synthetic jet. Due to the fact that fluctuations in the lift and drag coefficients on the non-actuated airfoil are characterized by vortex shedding to a certain extent, it is of interest that the effects of synthetic jet actuation on the spectra of lift and drag coefficients are investigated. Since the vortex shedding mode is usually identified by its corresponding spectral peak and such information essentially provides some insight into flow control, the objective of this particular set of study is to observe the effects of synthetic jet actuation on the fluctuating lift and drag coefficients on the airfoil. Since the fluctuating lift and drag coefficients are normally governed by the vortex shedding mode when the airfoil is not excited, it is of interest to see whether the characteristics of these fluctuations could be modified by the synthetic jet actuation provided that the momentum coefficient is sufficiently high.

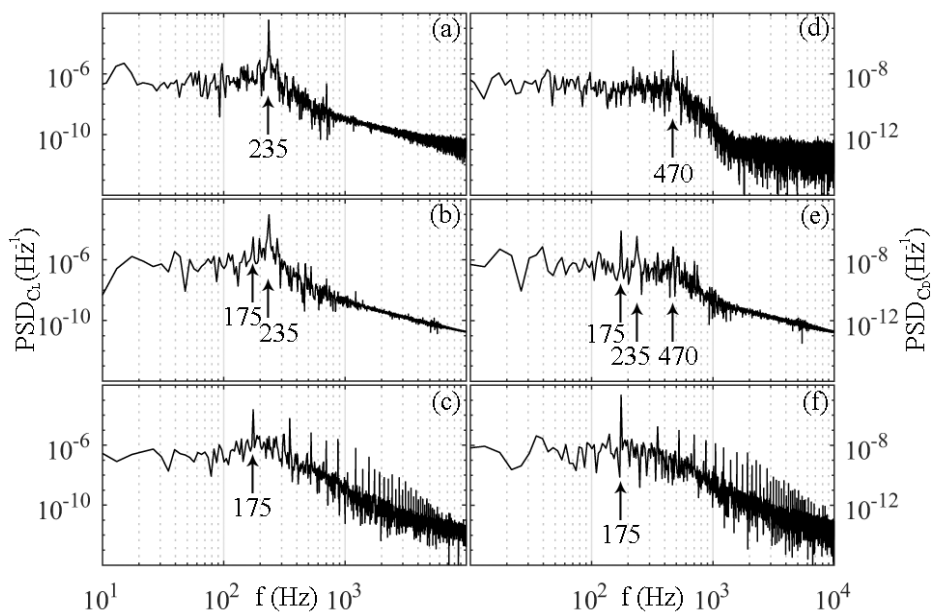

**Figure 9 Spectra of fluctuating lift coefficients on (a) the non-actuated airfoil and actuated airfoil when $C_\mu$ = (b) 0.0044 and (c) 0.0688 and spectra of fluctuating drag coefficients on (d) the non-actuated airfoil and actuated airfoil when $C_\mu$ = (e) 0.0044 and (f) 0.0688.**

In an attempt to answer the above question, spectra of fluctuating lift and drag coefficients on the non-actuated and actuated airfoils are shown in Figure 9 to provide some insight into the effects of synthetic jet actuation on the fluctuating lift and drag frequencies. For the sake of simplicity and conciseness, only the spectra from $\alpha = 0°$ are presented. As mentioned in the previous section, the fluctuating lift and drag frequencies on the non-actuated airfoil are characterized by the vortex shedding frequency (Figure 9a and d). Although a weak spectral peak at 175 Hz could be found on the lift spectra when a synthetic jet is actuated at $C_\mu = 0.0044$ (Figure 9b), it is noted that a spectral peak at 235 Hz which represents vortex shedding is still the dominant mode. Nevertheless, the magnitude of a spectral peak at 175 Hz on the drag spectra is observed to be higher than that of the fluctuating drag frequency which is normally twice the vortex shedding frequency (470 Hz). Additionally, the magnitude of a spectral peak at 235 Hz is observed to exceed that of a spectral peak at 470 Hz (Figure 9e). Due to the intervention of a synthetic jet, a reduction in the magnitude of a spectral peak at 470 Hz which is accompanied by an increase in the magnitude of a spectral peak at 235 Hz may occur as a result of imbalanced vorticities between the suction and pressure sides of the airfoil. When a synthetic jet is actuated at high momentum coefficient ($C_\mu = 0.0688$), the spectral peaks at 175 Hz become very apparent on both lift and drag spectra (Figure 9c and f). Further observation reveals that their subharmonics could also be found at every increment of 175 Hz. Unlike the cases of zero (non-actuated case) and low momentum coefficients whose spectra imply that their fluctuating aerodynamic coefficients are mostly and partly governed by the vortex shedding frequency, respectively, the lift and drag spectra at a high momentum coefficient suggest that their fluctuations may be dictated by the synthetic jet frequency. In order to further explore the likeliness of this possibility, time-history profiles of fluctuating lift and drag coefficients of the non-actuated and actuated airfoil ($C_\mu = 0.0044$ and 0.0688) are provided for the discussion.

Figure 10 and Figure 11 show the sequences of the synthetic jet actuation and the time-history profiles of fluctuating lift and drag coefficients when a synthetic jet is actuated at $C_\mu = 0.0044$ and 0.0688, respectively. Since the objective of this section is to attempt to understand the effects of synthetic jet actuation on fluctuating lift and drag coefficients, the time-history profiles of fluctuating aerodynamic coefficients on the actuated airfoil and synthetic jet velocity measured at the orifice are collected from the period of time when these parameters have already reached their quasi-periodic states or convergence. On the same graph, the time-history profiles of fluctuating aerodynamic coefficients on the non-actuated airfoil, which are obtained from a separate simulation, are plotted alongside the existing profiles to provide some comparisons between the waveform fluctuations of the non-actuated and actuated airfoils.

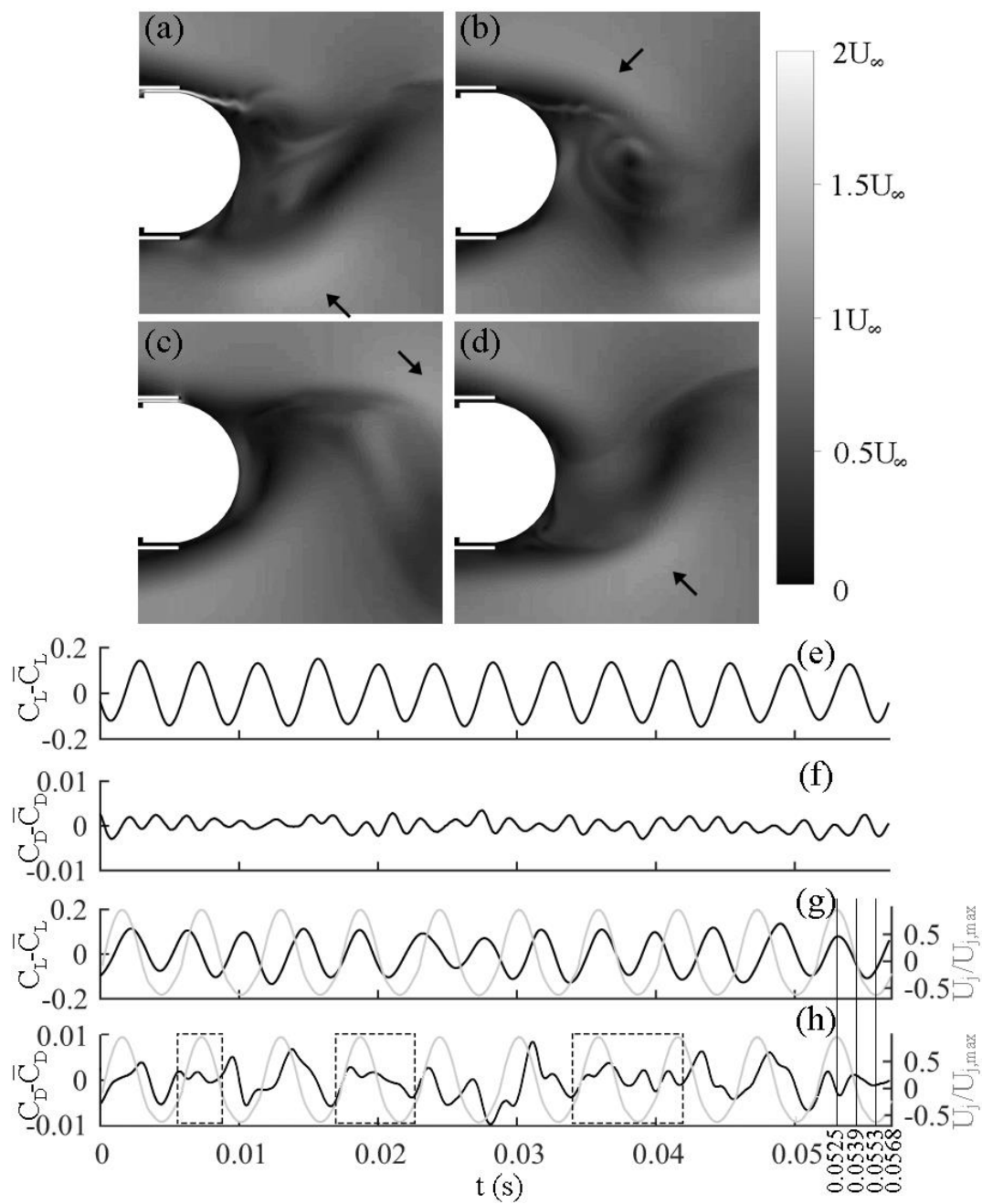

**Figure 10 Sequence of the synthetic jet actuation ($C_\mu = 0.0044$) as illustrated by contour maps of instantaneous velocity at $t/T_{act}$ = (a) 0.25, (b) 0.50, (c) 0.75, and (d) 1.00 and fluctuating (e) lift and (f) drag coefficients on the non-actuated aerofoil and fluctuating (g) lift and (h) drag coefficients (black trend) on the actuated aerofoil at $\alpha = 0°$ plotted against the synthetic jet velocity profile (grey trend).**

Figure 10a, b, c, and d are taken from $t/T_{act} = 0.25$ (expulsion stroke), 0.5, 0.75 (ingestion stroke), and 1.00 which are corresponding to $t =$ 0.0525, 0.0539, 0.0553, and 0.0568s in Figure 10g and h, respectively. According to the lift and drag spectra when $C_\mu = 0.0044$ (Figure 9c and d), these spectra seem to suggest that fluctuations in the lift coefficients are still mostly governed by the vortex shedding mode (235 Hz) whereas the drag coefficient is observed to fluctuate at mixed frequencies of a synthetic jet (175 Hz), vortex shedding (235 Hz), and twice the vortex shedding (470 Hz). Although the synthetic jet frequency is found to be the dominant mode in the drag spectrum, evidences on the existence of vortex shedding are still quite clear from the spectra (Figure 9b) and the flow visualization. As shown in Figure 10a-d, a trail of vortices could be seen in the wake of the airfoil and these vortices are indicated by the light contours and the arrows. Similar to fluctuating lift coefficients on the non-actuated airfoil, fluctuations in the lift coefficients at $C_\mu = 0.0044$ are observed to follow the sequences of vortex shedding in which the peaks of the lift coefficient, which are obtained within $0.0525 < t < 0.0539$s and at $t =$

0.0568s in Figure 10g, are corresponding to vortex shedding on the pressure side of the airfoil (Figure 10a and d) whereas the trough of the lift coefficient, which is obtained within $0.0539s < t < 0.0553s$ in Figure 10g, is corresponding to vortex shedding on the suction side of the airfoil (Figure 10b-c). Regarding the fluctuations in the drag coefficients at $C_\mu = 0.0044$, the time-history profile of drag coefficients shows that some cycles of fluctuating drag coefficients begin to fluctuate at the synthetic jet frequency (uncropped timeframe in Figure 10h) while some other cycles continue to fluctuate at frequencies higher than the SJ frequency frequency (cropped timeframe in Figure 10h).

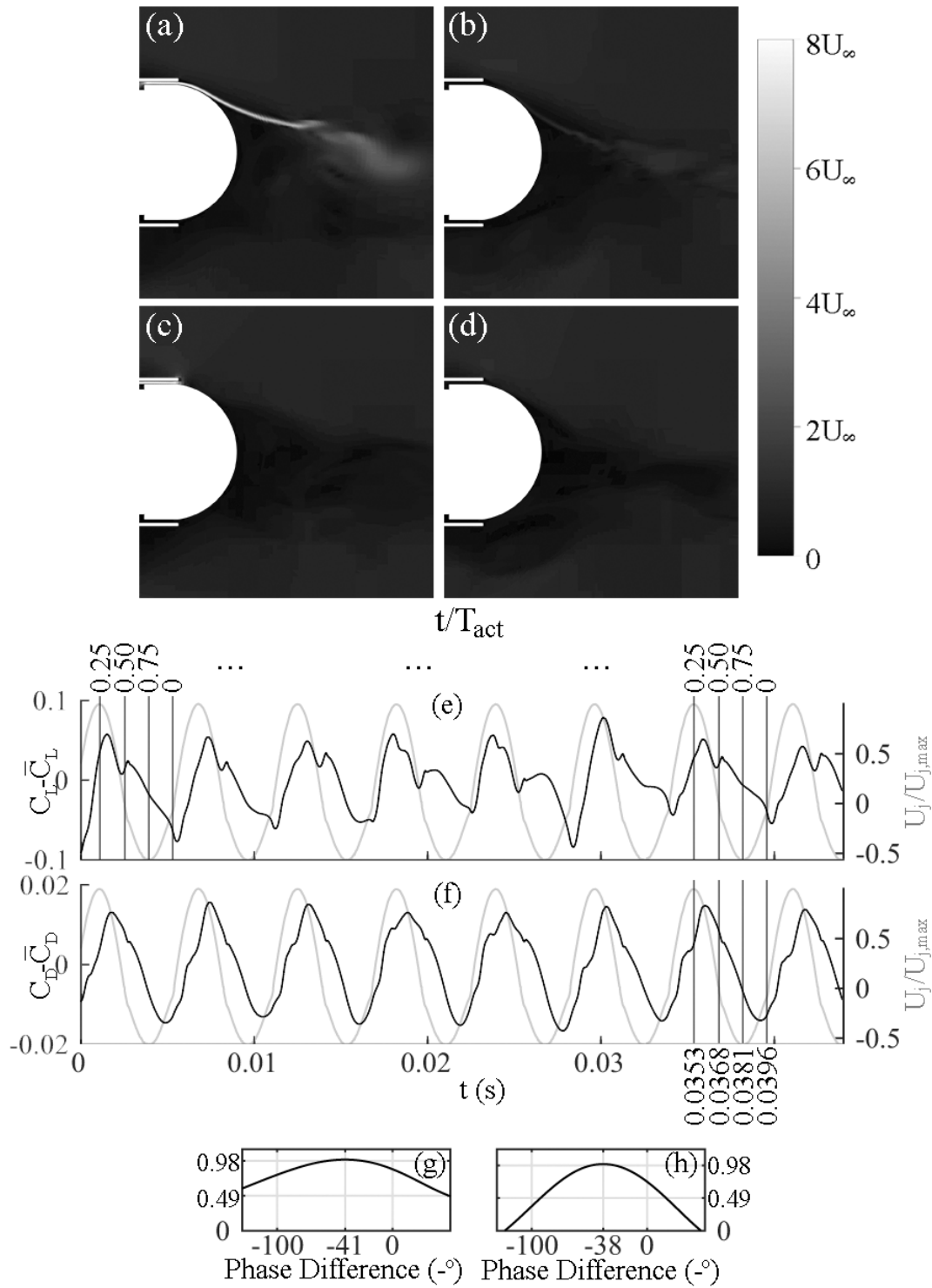

**Figure 11** Sequence of the synthetic jet actuation ($C_\mu = 0.0688$) as illustrated by contour maps of instantaneous velocity at $t/T_{act}$ = **(a)** 0, **(b)** 0.25, **(c)** 0.50, and **(d)** 0.75, series of fluctuating **(e)** lift and **(f)** drag coefficients (black trend) on the actuated aerofoil at α = 0° plotted against the synthetic jet velocity profile (grey trend), and cross-correlations of aerodynamic coefficients ((g) lift and (h) drag coefficients) and the synthetic jet velocity profile.

Figure 11a, b, c, and d are taken from $t/T_{act} = 0.25$ (expulsion stroke), 0.5, 0.75 (ingestion stroke), and 1.00 which are corresponding to $t = 0.0353, 0.0368, 0.0381,$ and $0.0396$s in Figure 11e and f, respectively. As opposed to the case of $C_\mu = 0.0044$, it is evident from the spectra (Figure 9c and f) and time-history profiles of both lift and drag coefficients that fluctuations in these aerodynamic coefficients (Figure 11e and f) are dictated by the synthetic jet actuation at $C_\mu = 0.0688$. In this case where the lift and drag profiles are strongly correlated to the synthetic jet profile, the cross-correlations of the aerodynamic (lift and drag) profiles and the synthetic jet profile (Figure 11g and h) reveal that the time series of the lift and drag profiles lag behind that of the synthetic jet profile by approximately 41° and 39° ($9.14 \times 10^{-4}$s), respectively. Since the lift and drag profiles at $C_\mu = 0.0688$ fluctuate at the synthetic jet frequency and their phase angles lag behind the synthetic jet profile by about 40° on the average, it is of interest that the characteristics of these consistent lift (Figure 11e) and drag (Figure 11f) waveforms are described with respect to the cycle of the synthetic jet actuation.

From $t/T_{act} = 0$ to 0.25, the trends of both lift and drag profiles are found to follow the uphill trip of the synthetic jet profile. Followed by the peak of the synthetic jet profile ($0.25 < t/T_{act} < 1.00$), the lift profile develops 2 peak values at $t/T_{act} \approx 0.30$ and 0.60 while the drag profile develops a single peak value at $t/T_{act} \approx 0.36$. Due to the fact that the synthetic jet actuation consists of successive blowing (Figure 11a) and suction (Figure 11c), 2 vortices which are produced from the orifice edge during blowing ($0 < t/T_{act} < 0.50$) and suction ($0.50 < t/T_{act} < 1.00$) are likely to be responsible for the double-peak lift profile. Followed by the peaks of both lift and drag profiles, the trend of the lift profile continues to drop during $0.60 < t/T_{act} < 1.00$ whereas that of the drag profile falls and rises as a synthetic jet undergoes its ingestion stroke at $0.50 < t/T_{act} < 0.75$ and returns to its neutral stage ($U_j/U_{j,max} = 0$) at $0.75 < t/T_{act} < 1.00$, respectively. The complete time-history profiles of these fluctuating lift and drag coefficients show that such fluctuations continue to repeat their cycles at the pace of a synthetic jet.

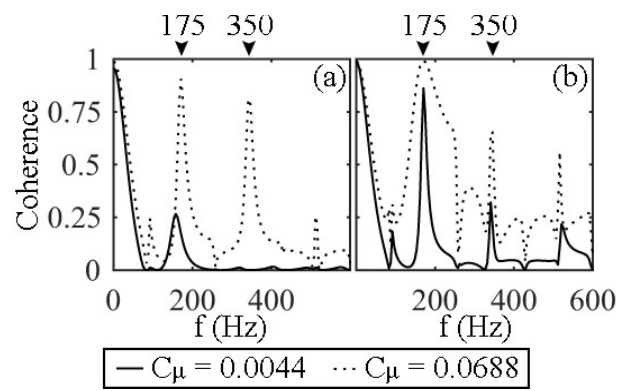

**Figure 12 Magnitude-squared coherence of series of fluctuating lift and drag coefficients to a series of synthetic jet velocity when a synthetic jet is actuated at $C_\mu$ = (a) 0.0044 and (b) 0.0688.**

As indicated by the lift and drag spectra shown in Figure 9, increments in the magnitudes of the spectral peaks at the synthetic jet frequency imply that the lift and drag profiles tend to fluctuate at the synthetic jet frequency as the momentum coefficient increases. Nevertheless, since these results only take the lift and drag profiles (output data) into account, the magnitude-squared coherence between the synthetic jet profile (input data) and the lift and drag profiles (output data), as shown in Figure 12, is computed so that the degree of similarly between the input and the output could be assessed. In the case of the lift profile, the coherence between the synthetic jet profile and the lift profile at 175 Hz is relatively weak at $C_\mu = 0.0044$ and is becoming stronger at $C_\mu = 0.0688$. In the case of the drag profile, strong coherence between the synthetic jet profile and the drag profile at 175 Hz has already been established by the time a synthetic jet is actuated at $C_\mu = 0.0044$. Exciting the airfoil at $C_\mu = 0.0688$ further increases the coherence between the 2 profiles. Unlike the waveforms of the lift profile at $C_\mu = 0.0688$ which exhibit double-peak profiles, it is highly possible that the coherence between the synthetic jet profile and the drag profile is higher than that of the lift profile due to the fact that the waveforms of the drag profile exhibit single-peak profiles which are much more similar to the waveforms of the synthetic jet profile.

### 4.3 Mean Lift Coefficient

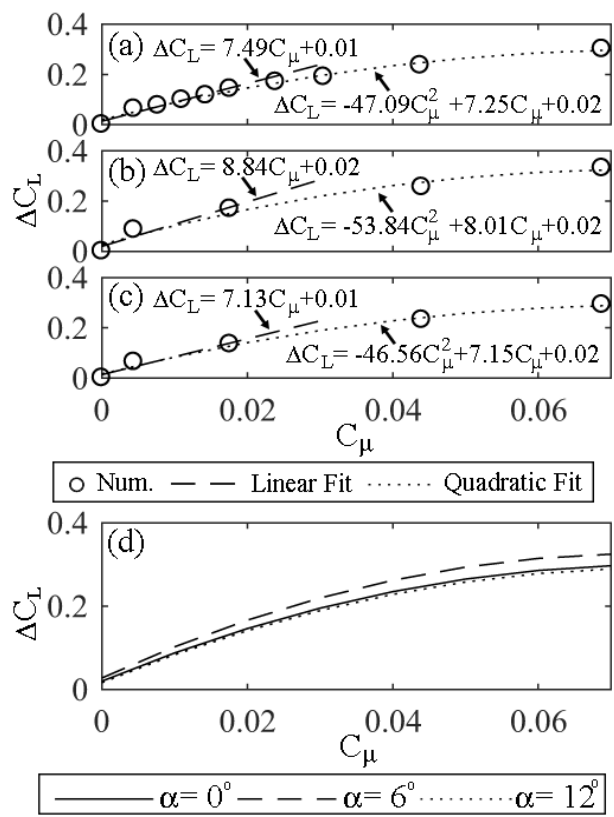

**Figure 13 Effects of momentum coefficient on the incremental lift coefficient of the airfoil at $\alpha$ = (a) 0°, (b) 6°, and (c) 12° and (d) their quadratic fits.**

Figure 13a-c show the effects of momentum coefficient on the lift increment slopes at $\alpha = 0°$, 6°, and 12° while Figure 13d compares the quadratic fits of the lift increment slopes at these angles of attack. In Figure 13a-c, dash lines and dot lines are drawn to provide linear fits to the plots at $0 < C_\mu < 0.0172$ and quadratic fits to the plots at $0 < C_\mu < 0.0688$, respectively. Although the linear fits are provided to the plots at $0 < C_\mu < 0.0172$, these linear fits are extended to $C_\mu = 0.0300$ so that the differences between the linear and quadratic fits at $C_\mu > 0.0172$ could be seen. Since the lift increment slopes at $0 < C_\mu < 0.0044$ which are estimated from the numerical study agree well with those acquired from the experimental study (See Figure 4), the lift increment slopes at $0.0044 < C_\mu < 0.0688$ are further estimated from the numerical approach to explore the effects of synthetic jet actuation at higher momentum coefficients. At $0 < C_\mu < 0.0172$, the lift coefficient is observed to almost linearly increase with the momentum coefficient. Although the lift coefficient continues to increase at $0.0172 < C_\mu < 0.0688$, the rates of lift increment gradually decline as the synthetic jet actuation is approaching $C_\mu = 0.0688$. Due to these gradual declines in the rates of lift increment at a high momentum coefficient range, a quadratic function has the best fit to these lift increment slopes. Furthermore, the quadratic fits of the lift increment slopes suggest that the effectiveness of the synthetic jet actuation may be somewhat dependent on the angle of attack as the lift increment slope slightly increases and decreases at $\alpha = 6°$ and 12°, respectively. In the experimental study [15], the standard deviations of fluctuating surface pressures suggest that the effectiveness of the synthetic jet actuation may be related to a boundary layer state as the synthetic jet actuation on the high-incidence airfoil is observed to yield lower lift increment slope. The cause of the reduced effectiveness may be due to an increase in the turbulence level near the synthetic jet slot which could be attributed to the breakdown of vortical structures on the airfoil.

## 5 Conclusions

In this study, the current approach to model the circulation control airfoil using synthetic jets was found to yield satisfactory results. The differences between the experimental and numerical results are typically 1-12% for lift coefficient. The differences between the locations of the flow separation and reattachment points obtained from the experimental and numerical studies are within 1-6%. The differences between the vortex shedding frequencies obtained from the experimental and numerical studies are less than 12%. According to these spectra, it was found that the fluctuating lift frequency is identical or similar to the vortex shedding frequency in some cases. Interestingly, the fluctuating drag frequency of the circulation control airfoil at $\alpha = 0°$ exhibits great similarity to that of a circular cylinder in which the drag coefficient was found to fluctuate at twice the vortex shedding frequency. Although a weak spectral peak at about twice the vortex shedding frequency was also found on the drag spectra at $\alpha = 6°$, its magnitude was so low that a "hump" was formed instead of a well-defined spectral peak. According to the drag spectra at $\alpha = 6°$, it would seem that the magnitude of a spectral peak at about twice the vortex shedding frequency decreases as the angle of attack increases. At $\alpha = 12°$, its fluctuating drag frequency is identical to its fluctuating lift frequency. Upon actuating a synthetic jet on the airfoil, it was found that the fluctuating lift and drag coefficients tend to fluctuate at the same frequency as a synthetic jet as the momentum coefficient increases. The lift coefficient was found to linearly increase with the momentum coefficient at a low momentum coefficient range. Nevertheless, a further observation on the relationship between the lift coefficient and momentum coefficient reveals that a quadratic function has the best fit for this plot as the incremental rate of the lift coefficient begins to decline at a high momentum coefficient range $(0.0172 < C_\mu < 0.0688)$. The quadratic fits of the lift increment slopes suggest that the effectiveness of the synthetic jet actuation may be somewhat dependent on the angle of attack as the lift increment slope slightly increases and decreases at $\alpha = 6°$ and $12°$, respectively.

# Acknowledgments

The authors gratefully acknowledge the contribution of NeSI high-performance computing facilities to the results of this research. New Zealand's national facilities are provided by the New Zealand eScience Infrastructure (NeSI) and funded jointly by NeSI's collaborator institutions and through the Ministry of Business, Innovation and Employment. URL: www.nesi.org.nz